# Down with ncRNA! Long live fRNA and jRNA!

Dan Graur

Department of Biology & Biochemistry
University of Houston
Science & Research Building 2
3455 Cullen Blvd. Suite #342
Houston, TX 77204-5001

Voice: 713-743-7236
Fax: 713-743-2636
Email: dgraur@uh.edu

Abstract

Noncoding RNA (ncRNA) and long noncoding RNA (lncRNA) are scientifically invalid terms because they define molecular entities according to properties they do not possess and functions they do not perform. Here, I suggest retiring these two terms. Instead, I suggest using an evolutionary classification of genomic function, in which every RNA molecule is classified as either "functional" or "junk" according to its selected effect function. Dealing with RNA molecules whose functional status is unknown require us to phrase Popperian nomenclatures that spell out the conditions for their own refutation. Thus, in the absence of falsifying evidence, RNA molecules of unknown function must be considered junk RNA (jRNA).

Negative descriptions in biology are generally considered invalid. That is, biological entities cannot be solely defined by what they do not possess or do not do. Hence, for instance, the taxon Pisces (fishes) has been deemed scientifically invalid even before its monophyletic status was refuted, because the definition of Pisces involved a single negative character state—the lack of limbs with digits. The same principles should apply to the taxonomy of molecular entities.

In the scientific literature, the modifiers "non-coding," "noncoding," and "nc" are widely used as prefixes for "DNA" and "RNA." As of September 1, 2017, these terms appear more than 45,000 times in Google Scholar. This negative description of RNA and DNA is particularly troublesome because there literally exist an infinite number of functions that ncRNA does not perform. Thus, one might as well call such molecules non-dancing DNA (ndDNA) or non-triple-jumping RNA (ntjRNA).

The ncRNA category lumps together a heterogeneous collection of biological entities that have either unrelated functions or have no function, and which exhibit no sequence similarity to one another. Thus, neither functional nor evolutionary justification exists for this catchall category. To make matters worse, the division of the genome into coding and noncoding gives rise to an unfortunate phenomenon, whereby whenever a function is found for a particular RNA molecule, the finding is interpreted as a refutation of the junk DNA concept (e.g., Wright and Bruford 2011; Pennisi 2012; Francis Collins quoted in Zimmer 2015; Younger and Rinn 2015; Saey 2016). Google Scholar searches reveal that phrases such as "once thought to be junk DNA," and "once considered to be junk DNA" appear in the scientific literature in excess of 350 times. The number of times such phrases appear in the nonscientific media is immeasurably greater.

Here, I suggest retiring all meaningless terms with the modifier "non-coding." Instead, I suggest using the evolutionary classification of genomic function suggested by Graur et al. (2015), in which every genomic segment and every genomic product (RNA or protein) that do not decrease the fitness of their carriers is classified as either "functional" or "junk." Thus, every RNA molecule that has a selected effect function, i.e., every RNA molecule whose existence is due to the function for which it was selected, should be referred to as functional RNA (fRNA). All other RNA molecules should be classified as junkRNA (jRNA). Table 1 lists some examples of fRNA.

As opposed to ncRNA, the term long noncoding RNA (lncRNA) is only partially meaningless because in addition to its being defined negatively, the definition also includes a positive, albeit arbitrary character. To be called lncRNA, an ncRNA molecule must be 200 nucleotides or longer. Unfortunately, while not being meaningless *per se*, this term is useless. In systematics, the term "garbage-can taxon" or "rubbish-bin taxon" refers to a catchall category, into which evolutionarily unrelated taxa are dumped for the sole reason that they do not fit into any other existing taxonomic category. One such example is Insectivora, which is defined positively by diet, yet is paraphyletic. Analogously, lncRNA is a "garbage can" category, into which different types of RNA are dumped. Many are functionless, while others have been shown to have widely disparate roles that are unrelated to one another. There is no reason to place all fRNAs larger than 200 nucleotides into a single category; instead functional lncRNAs should be divided into different classes by function or evolutionary relatedness, and the rest should be added to the jRNA category.

Dealing with RNA molecules that may or may not have a function seems at first difficult. In practice, however, such cases should be treated in a manner that is respectful of the Popperian dictum, according to which scientific hypotheses (and nomenclatures) should be phrased in such a manner as to spell out the conditions for their own refutation (Graur 2016). Should we, in the absence of evidence, assume functionality or lack of functionality? Let us consider both cases. A statement to the effect that an RNA molecule is devoid of a selected-effect function can be easily rejected by showing that the element evolves in a manner that is inconsistent with neutrality. If, on the other hand, one assumes as the null hypothesis that the RNA is functional, then failing to find telltale indicators of selection cannot be interpreted as a rejection of the hypothesis, but merely as a sign that we have not searched thoroughly enough or that the telltale signs of selection have been erased by subsequent evolutionary events. There exists a fundamental asymmetry between verifiability and falsifiability in science: scientific hypotheses can never be proven right; they can only be proven wrong. The hypothesis that a certain genomic element is functional can never be rejected and is, hence, unscientific. According to physicist Wolfgang Pauli (quoted in Peierls 1960), a hypothesis that cannot be refuted "is not only not right, it is not even wrong."

Literature

Graur D. 2016. *Molecular and Genome Evolution*. Sinauer Associates, Sunderland, MA

Graur D, Zheng Y, Price N, Azevedo RB, Zufall RA, Elhaik E. 2013. On the immortality of television sets: "function" in the human genome according to the evolution-free gospel of ENCODE. Genome Biol. Evol. 5:578-590.

Graur D, Zheng Y, Azevedo RB. 2015. An evolutionary classification of genomic function. Genome Biol. Evol. 7:642-645.

Iyer MK, et al. 2015. The landscape of long noncoding RNAs in the human transcriptome. Nat. Genet. 47:199-208.

Peierls, R. 1960. Wolfgang Ernst Pauli, 1900–1958. Biogr. Mem. Fellows R. Soc. 5: 174–192.

Pennisi E. 2012. ENCODE Project writes eulogy for junk DNA. Science 337:1159-1161

Saey TH. 2016. 'Junk DNA' has value for roundworms. Science News https://www.sciencenews.org/article/%E2%80%98junk-dna%E2%80%99-has-value-roundworms

Wright MW, Bruford EA. 2011. Naming 'junk': Human non-protein coding RNA (ncRNA) gene nomenclature. Hum. Genomics 5:90-98.

Younger ST, Rinn JL. 2015. Silent pericentromeric repeats speak out. Proc. Natl. Acad. Sci. USA 112:15008-15009.

Zimmer C. 2015. Is most of our DNA garbage? New York Times http://www.nytimes.com/2015/03/08/magazine/is-most-of-our-dna-garbage.html

Table 1. Examples of fRNAs

---

| Class | Abbreviation | Function |
|---|---|---|
| Guide RNA | gRNA | Template for posttranscriptional RNA editing |
| Micro RNAs | miRNA | Posttranscriptional and translational regulation |
| Ribosomal RNA | rRNA | A component of the ribosome |
| Small interfering RNA | siRNA | RNA interference |
| Piwi-interacting RNA | piRNA | Transcriptional silencing of retrotransposons |
| Small nuclear RNA | snRNA | Splicing of spliceosomal introns |
| Small nucleolar RNA | snoRNA | Guide methylation/pseudouridylation of RNA, removal of introns from pre-mRNA, regulation of transcription factors and RNA polymerase II, maintaining telomeres |
| Small temporal RNA | stRNA | Regulate gene expression by preventing the mRNAs they bind to from being translated |
| Transfer RNA | tRNA | Adaptor molecule for amino acids in protein synthesis |

| Ribozyme | | Catalysis of chemical reactions |
|---|---|---|
| Transfer-messenger RNA | tmRNA | Rescuing stalled ribosomes, tagging for degradation incomplete polypeptide chains, promoting degradation of aberrant mRNA |
| RNA in ribonucleoproteins | | Component of RNA-protein functional complexes |